# From Filter Paper to Functional Actuator by Poly(ionic liquid)-Modified Graphene Oxide

*Haojie Song,[1,2] Huijuan Lin,[1] Markus Antonietti,[1] and Jiayin Yuan[1]\**

[1]Prof. H. Song,[+] H. Lin,[+] Prof. M. Antonietti, Dr. J. Yuan
Department of Colloid Chemistry, Max Planck Institute of Colloids and Interfaces,
D-14424 Potsdam, Germany
E-Mail: jiayin.yuan@mpikg.mpg.de
[2]School of Materials Science and Engineering, Jiangsu University, Zhenjiang, Jiangsu, 212013, China
[+] These authors contributed equally to this work.

Abstract

A commercially available membrane filter paper composed of mixed cellulose esters bearing typically an interconnected pore structure was transformed into a stimuli-responsive bilayer actuator by depositing a thin film of poly(ionic liquid)-modified graphene oxide sheets (GO-PIL) onto the filter paper. In acetone vapor, the as-synthesized bilayer actuator bent readily into multiple loops at a fast speed with the GO-PIL top film inwards. Upon pulling back into air the actuator recovered their original shape. The asymmetric swelling of the top GO-PIL film and the bottom porous filter paper towards organic vapor offers a favorably synergetic function to drive the actuation. The PIL polymer chains in the hybrid film were proven crucial to enhance the adhesion strength between the GO sheets and the adjacent filter paper to avoid interfacial delamination and thus improve force transfer. The overall construction allows a prolonged lifetime of the bilayer actuator under constant operation, especially when compared to that of the GO/filter paper bilayer actuator without PIL**.**

# 1. Introduction

Stimuli-responsive materials and devices, which exhibit large shape deformation or color alternation in response to variation in external environment, such as temperature, pH, solvents, ionic strength, electromagnetic field, humidity and the redox state, are receiving widespread interest.[1-8] The pursuit and development of novel, emerging functional materials provides rich opportunities for the design and construction of task-specific actuators. Additionally, driven by the desire for material sustainability, it is a tendency to fabricate actuators at least partially from biopolymers or bio-templates.[9,10] For example, Whitesides reported the development of soft pneumatic actuators based on composites consisting of elastomers with embedded papers that are flexible but inextensible.[10]

In parallel, porous polymer actuators have emerged as a new design methodology, inspired by the cellular structures found in natural plants, such as conifer cones and ice plants.[11] For example, porous polymer membranes built up from poly(ionic liquid)s (PILs),[12] were developed into a multi-responsive, ultrafast and vapor-driven light-weight actuator. In that synthesis, an aqueous solution of ammonia was applied to a PIL-based polymer blend film on a glass plate, and the diffusion of the ammonia molecules induced a porous architecture with a gradient structure inside the blend film. The as-synthesized porous membrane uniquely was possessed of multi-responsiveness towards a variety of organic vapors in both the dry and wet state. Similarly, an asymmetric, porous, self-folding polyimide actuator was fabricated by using silica particles mixed with a polyimide precursor solution, which after particle sedimentation and formation of polyimide were etched away to engineer the pores inside the bottom zone of the polyimide film.[13] The driving force for actuation here was created from the different swelling of porous and nonporous parts of the same film. The introduction of porosity into the materials in general is promising for the development of high performance actuators, as the pores not only make the actuator light-weight, but also accelerate the mass transport and diffusion rate to realize faster motion.

Graphene as such provides large surface area, superior carrier transport, mechanical bendability, and thermal/chemical stability, and is non-swellable, but shows great potential as basic building blocks for advanced actuators.[14-20] Recently, rapid progress has been made in graphene based electromechanical actuators.[21-23] Graphene, graphene oxide (GO) and graphene based composite materials have been used to convert electrical energy to mechanical energy. For example, Dai *et al.* have recently developed a unimorph electrochemical actuator based on a monolithic graphene film with asymmetrically modified surfaces.[24] In addition to electrical and electrochemical stimulus, a macroscopic assembly of GO and multi-walled CNT (MWCNT) bilayer paper demonstrated interesting humidity and/or temperature dependent actuation behavior.[25,26] GOs were however previously not applied in organic vapor driven actuators, because of their inertness and the gas barrier nature of GO nanosheets towards organic vapor.

Herein, we present simple construction of a solvent vapor-driven actuator based on a bilayer membrane configuration, which consists of a filter paper composed of mixed cellulose esters as a stiff and stable porous substrate and a top coating layer from poly(ionic liquid)-modified GO nanosheets (GO-PIL). The different response of the substrate and the top coating layer towards organic vapor enables the bilayer membrane to undergo fast, reversible bending/unbending motions. Our data support that the surface decoration of GOs by PILs is decisive. Compared to a coating layer built up exclusively from pristine GOs, the PIL modified GO layer can efficiently extend the lifetime of such bilayer actuators by providing strong binding power between GOs and the filter paper surface to avoid adhesion failure and delamination.

## 2. Results and Discussion

The fabrication of the bilayer membrane actuator from PIL-grafted GO nanosheets (GO-PIL) and the filter paper is illustrated in **Figure 1**. The strong electrostatic interactions between negatively charged GO nanosheets and cationic PIL polymer chains have been previously reported to stabilize or chemically reduce GO nanosheets in a solution state.[27] This feature was exploited in this study to attach PIL chains onto the GO nanosheet surface, which is proven by the Zeta potential measurements with -38.3 mV in pure GO aqueous solution and only -0.3 mV in GO-PIL aqueous solution, indicating the negative charge on GO nanosheets was nearly neutralized by the positively charged PIL chains. Different from a simple physical mixture of GO nanosheets and PIL in a solvent that gives immediate nondispersible aggregates, the present stabilization process was achieved simultaneously by radical polymerization of the IL monomer, in which the *in situ* generated PIL polymer chains were electrostatically absorbed onto the surface of GO nanosheets. This stabilization method has been previously proven to be powerful to disperse negatively charged nanomaterials in a liquid phase.[28] Upon filtration through a filter paper with three-dimensionally interconnected submicron pores, the micrometer-sized GO-PIL nanosheets were deposited flatly onto the top of the filter paper to build up a thin coating layer. Through elemental analysis of the hybrid GO-PIL nanosheets to determine the nitrogen content, PIL stands for 25.9 wt.% of the weight of the as-synthetized GO-PIL film, and 14.3 wt.% of the overall hybrid bilayer membrane.

The existence of the PIL polymer chains in the dried GO nanosheets powder was further verified by analyzing the FT-IR spectra of the pristine GO powder, pure PIL and GO-PIL in **Figure 2**a. The obvious IR bands at 3422, 1746, 1620, and 1055 $cm^{-1}$ in the FT-IR spectrum of GO were attributed to O−H stretching, C=O carbonyl stretching, aromatic C=C stretching, and symmetric C−O stretching in the C−O−C groups, respectively. After modification with PIL, new bands at 2918, and 2846 $cm^{-1}$ in the high-frequency region appeared, which were typical for asymmetric and symmetric vibration of C−H stretching, respectively. Furthermore, the bands at 1160 and 1551 cm$^{-1}$ in the hybrid were assigned to C–N stretching of imidazolium rings and ring in-plane asymmetric stretching *CH$_2$*(N) as well as *CH$_3$*(N)CN stretching vibrations of the PIL side chains. According to the FT-IR results, it was concluded that the surfaces of GO were indeed grafted with PILs, which is in good agreement with the results of the Zeta potential measurements. Thermo-gravimetric analysis was conducted to determine the thermal stability of GO before and after PIL functionalization in Figure 2b. It can be seen that while PIL starts an obvious decomposition step at 200°C, GO starts to lose mass only slowly upon heating up to 300 °C, which can be attributed to the loss of molecularly adsorbed water and the dehydration process. The major mass loss of GO occurs right at 300 °C, presumably due to pyrolysis of the labile oxygen-contained functional groups to yield CO or $CO_2$. Similar to GO, a weight loss of approximately 10 % is observed at 200 °C for GO-PIL. However, the trend of thermal degradation of GO-PIL above 200 °C is different from that of the GO. Notably, that rapid thermal decomposition of GO-PILs takes place at ~250 °C due to the thermal degradation of PILs on the surface of GO. This may be caused by the well-known retro-quaternization reaction in the PIL.

The microstructure of PIL modified GO nanosheets was examined by transmission electron microscopy (TEM). TEM observation in Figure 2c shows that GO nanosheets are highly transparent under electron beam with folding at edges, suggesting their nature of thin thickness. After modification of GO by PIL, the TEM image (Figure 2d) clearly illustrates that two coexisting phases, *i.e.* GO as light plate and PIL chain aggregates as dark dots, are observed and mixed at a nanoscale to form an intimate interface. Such a unique inhomogeneous growth of polymer chains onto GO nanosheets *via in-situ* polymerization method has been reported previously.[29] The adhesion of the PIL onto the GO surface is robust, and even a long mechanical sonication treatment (full power for 30 min in an aqueous sonication bath, 140 watt) before the TEM characterization will not affect the mushroom-like morphology of the PIL chains on the surface of GO. These characterization results suggest that the PILs exist in the hybrid GO-PIL structure in a heterogeneous manner, *i.e.* neither in a mere physical mixture of two separate phases of GO and PIL, nor in a homogenous distribution of the PIL in the GO matrix that one would expect.

The structure of the as-prepared GO-PIL/filter paper bilayer membrane is revealed in **Figure 3**a. The cross-sectional image by scanning electron microscopy (SEM) in Figure 3b shows unambiguously a bilayer configuration of the GO-PIL/filter paper membrane (boundary indicated by a dashed yellow line). A top layer of assembled GO-PIL nanosheets stacking onto each other is visible, while the bottom filter paper with an interconnected porous network is tightly adhered to the GO-PIL top film (Figure 3c). The thickness of the GO-PIL film is controllable in terms of the used volume of the dispersion and is typical in the range of 1-10 μm, here of ca. 7 *μ*m in Figure 3c. The strong binding between the filter paper and the adjacent GO-PIL film benefits from the well-known adhesion power of PILs towards various types of surface, such as metals, metal oxides, cellulose and carbon nanostructures.[30] The filter paper used here shows a random distribution of interconnected sub-micron pores with a diameter in the range from 0.3 *μ*m to 1 *μ*m (Figure 3d). These interconnecting holes among neighboring macropores form a porous structure allowing for rapid vapor transport channel in the bilayer membrane. As expected, the GO-PIL film presents a uniformly smooth surface characteristic of the conventionally filtration-formed dense GO film (Figure 3e). Due to the inertness and gas barrier nature of GO nanosheets towards gas molecules, the GO-PIL film is of weak interaction with gas vapors.

The response of the as-synthesized GO-PIL/filter paper bilayer membrane was first investigated in an acetone vapor. When placed above a liquid acetone phase, the bilayer membrane with a length of 20 mm, width of 1 mm and thickness of 107 μm bent quickly into a closed loop in the first 1 s with the top GO-PIL layer inwards, and continued to bend further into a multiply wound coil (**Figure 4**a and Movie S1). Upon exposure back to air, the membrane recovered its original shape in 35 s due to the diffusion of acetone molecules out of the porous filter paper. The kinetics of the bending and unbending movements is assessed by plotting the curvature of the membrane against time (Figure 4b), showing a bending curvature of 0.53 mm$^{-1}$ in 5 s, that is, the 20 mm long membrane actuator closed one complete circle in about every 2s in average. At room temperature, the bending/unbending behavior of the GO-PIL/filter paper bilayer membrane toward periodic contact with acetone vapor was reversible and reproducible (Figure 4c). No detectable fatigue of the actuator was observed over time. In contrast, the filter paper alone only slightly deforms its shape when in contact with acetone vapor (Movie S2), because the stress on both sides of the filter paper surfaces generated by acetone sorption and the consequent symmetric swelling essentially offsets each other. In the GO-PIL/filter paper bilayer membrane, the asymmetry swelling in the actuator configuration was built up, as the bottom porous filter paper absorbed significantly more acetone molecules than the top dense GO-PIL layer, thus leading to the bending of the membrane.

In **Figure 5**a, it is further shown that the amplitude of locomotion of actuators of individual sizes in general is specific to the solvent, which is dictated by the solvent-filter paper interaction. For example, acetone and THF can disintegrate the filter paper, a measure of a strong solvent-cellulose ester interaction, and these solvents trigger the strongest actuation in terms of kinetics. Solvents like methanol and ethanol that only partly swell the filter paper drive the actuator with slower kinetics and weaker bending. On the contrary, solvents of low polarity that cannot swell the filter paper, such as hexane and toluene, fail to drive the actuator. Moreover, in the solvent of low vapor pressure at room temperature, such as DMSO, although it can dissolve the filter paper, indicative of strong interaction, the bilayer membrane undergoes little to no action in their vapor. Practically, the distinctive responsiveness towards different solvents may offer an alternative way to distinguish the type of solvents without complicated instrumental analysis.

The stability and reusability is a prerequisite for the actuators to be used in practical applications. Therefore, the curvature of the GO-PIL/filter paper bilayer membrane and GO/filter paper bilayer membrane in acetone vapor was studied in a reversible cycling test at the same condition. The curving of the GO/filter paper bilayer membrane and that of the GO-PIL/filter paper bilayer membrane are close at the initial stage of the cycling test (Figure 5b), say in the first 7 cycles. Right after that, the GO/filter paper bilayer membrane failed to operate reliably. In comparison, the GO-PIL/filter paper bilayer membrane keeps its bending performance almost unchanged upon extended cycling tests, here after 50 runs.

The characteristics of the interfacial region between GO film and filter paper, that is, the physico-chemical interactions at GO and filter paper interface, play an important role in the reusability of the GO/filter paper bilayer membrane. As shown in **Figure 6** (a, b, and c), in the absence of PILs, the top GO film after 7 cycles is attached poorly to the filter paper due to adhesion failure and delamination, a problem that is commonly observed in bilayer actuators. When GO nanosheets were modified by the PIL, the positively charged polymer chains on the surface of GO can serve as "glue", to enhance the adhesion between the GO-PIL and the filter paper substantially due to the non-convalent interaction (majorly hydrogen bonds and dipole-dipole interaction) between the imidazolium cations in polymer chains and oxygen-rich groups in filter paper. As shown in Figure 6 (d and e), the GO-PIL film sticks compactly to the filter paper, and the boundary of the GO-PIL film and the filter paper became blurry, which does not change after 50 runs.

Traditionally, asymmetric swelling of the polymer networks is the dominant mechanism to modulate hydrogel and ion gel actuators, the process of which is relatively slow as limited by the retarded diffusion of trigger molecules within the wet gels.[31] In the present actuator design, the bottom filter paper in its porous state allows the solvent molecules in a gas state to diffuse rapidly into the structure, while the top GO-PIL film with layered GO nanosheets densely packed onto each other blocks the diffusion of acetone vapor from one side and interacts poorly with acetone molecules (**Figure 7**). As such, the swelling of the hybrid membrane in the filter paper substrate bends the membrane actuator in a way that the porous filter paper part is outwards and the GO-PIL film inwards. In the degassing step, acetone molecules diffuse easily out through the macropores, which leads to the recovery of the original shape of the bilayer membrane.

## 3. Conclusion

A rationally designed porous bilayer actuator based on a poly(ionic liquid)-modified GO top film and a bottom filter paper that is commercially available is reported. In this unique example, the use of the reverse swelling and shrinking of the filter paper due to sorption and desorption of organic vapor leads to an excellent actuation response. By combination of GO-PIL nanosheets and porous filter paper with their orthogonal properties, a new actuator principle could be realized which is based on simple and well known materials, especially considering the broad variability of commercially available filter papers.

## 4. Experimental Section

*Materials:* 2,2'-Azobis(2-methylpropionitrile) (AIBN, 98%) was obtained from Sigma-Aldrich and recrystallized from methanol before use. 1-Cyanomethy-3-vinylimidazolium bromide (CMVImBr) was synthesized according to our previous report.[32] Graphene oxide (GO) (average diameter: 2.6 $\mu$m, thickness: 0.8-1.2 nm, and purity > 95%) was purchased from JCNANO company and used without further purification. The filter paper was composed of mixtures of cellulose acetate and cellulose nitrate (90 mm in diameter, 100 $\mu$m in thickness) provided by Merck Millipore. All solvents were of analytic grades.

*Preparation of GO-PIL/filter paper bilayer membranes*

*Preparation of the GO-PIL hybrid:* The GO nanosheets functionalized by the PIL poly(CMVImBr) were obtained by *in situ* polymerization of an ionic liquid monomer, 1-cyanomethyl-3-vinylimidazolium bromide (CMVImBr) in the presence of a dispersion of negatively charged GO in DMSO. During the polymerization, the *in situ* synthesized cationic PIL chains were attached to the GO surface *via* electrostatic interaction. In detail, first 50 mg of a GO nanosheet powder was sonicated in 50 ml of DMSO to form an optically homogeneous dispersion. It was followed by adding 0.5 g of the IL monomer and the mixture was stirred for 30 min. Then the dispersion was purged by dry nitrogen for another 30 min, followed by addition of 100 mg of AIBN as initiator to polymerize the IL monomer at 80 °C for 48 h. When cooling down to room temperature, the reaction mixture was dropwise added to an excess of THF. The dark precipitate was collected by filtration, re-dissolved in methanol and precipitated in THF again. The remaining black solid was dried under vacuum to produce PIL modified GO nanosheets (GO-PIL).

*Preparation of GO-PIL/filter paper bilayer membranes:* To prepare the GO-PIL/filter paper bilayer membrane, 10 mg mL$^{-1}$ of GO-PIL in water was sonicated to produce homogeneous dispersion. The bilayer membrane was fabricated *via* applying vacuum filtration of the GO-PIL dispersion in water through the membrane filter paper with an effective particle cutoff size of 0.05 $\mu$m. The resulting bilayer membrane was further rinsed with 100 ml of ultrapure Milli-Q water for 3 times. After the filtration, the membrane was dried for 24 h at 30 °C till constant weight. As a control experiment, GO nanosheets WITHOUT PIL-modification were used to produce GO/filter paper bilayer membranes in a procedure similar to that of GO-PIL/filter paper bilayer membrane.

*Characterization:* Fourier transform infrared (FT-IR) spectroscopy was performed at room temperature using a BioRad 6000 FT-IR spectrometer equipped with a Single Reflection Diamond ATR. Combustion elemental analysis was performed for carbon, hydrogen and nitrogen using a Vario MICRO Elementar. Thermogravimetric analysis (TGA) experiments were performed under nitrogen flow at a heating rate of 10 K min$^{-1}$ using a Netzsch TG209-F1 apparatus. The microstructures of the GO nanosheets before and after modification by PIL were examined by transmission electron microscopy (TEM) using an EM 912 Omega microscope at 120 kV. Morphologies of the membranes were examined by scanning electron microscopy (SEM) performed on a GEMINI LEO 1550 microscope at 3 kV, samples were sputtered with a thin layer of gold beforehand. For the zeta potential measurements, the samples of GO and GO-PIL after ultrasonication in water were separately loaded in folded capillary zeta potential cells. Three measurements for each sample were performed consisting of 20 runs with duration of 10 s.

*Solvent vapor stimulus actuation:* Organic solvents were loaded in a 100 ml glass beaker at 20 °C. Then a piece of bilayer membrane strip (20 mm ×1 mm × 107 $\mu$m) was placed 5 mm above the liquid phase of the solvent, where the solvent vapor will trigger the bending movement. Afterwards, the bilayer membrane was pulled back into air to accomplish the shape recovery. The bending curvature was calculated based on the formula of *c=1/r*, where *r* is radius of the bending loop.

**Supporting Information**
Supporting Information is available from the Wiley Online Library or from the author.

**Acknowledgements**
H. Song and H. Lin contributed equally to this work. We greatly appreciate the support from the Max Planck Society, the International Max Planck Research School (IMPRS) on ''Multiscale Biosystems'' and the National Natural Science Foundation of China (No. 51372103), the financial support from China Scholarship Council (201308320062), and the discussion with Dr. John W. C. Dunlop about the actuation mechanism. This research is partially supported by the ERC (European Research Council) Starting Grant with project number 639720 – NAPOLI.

Received: ((will be filled in by the editorial staff))
Revised: ((will be filled in by the editorial staff))
Published online: ((will be filled in by the editorial staff))

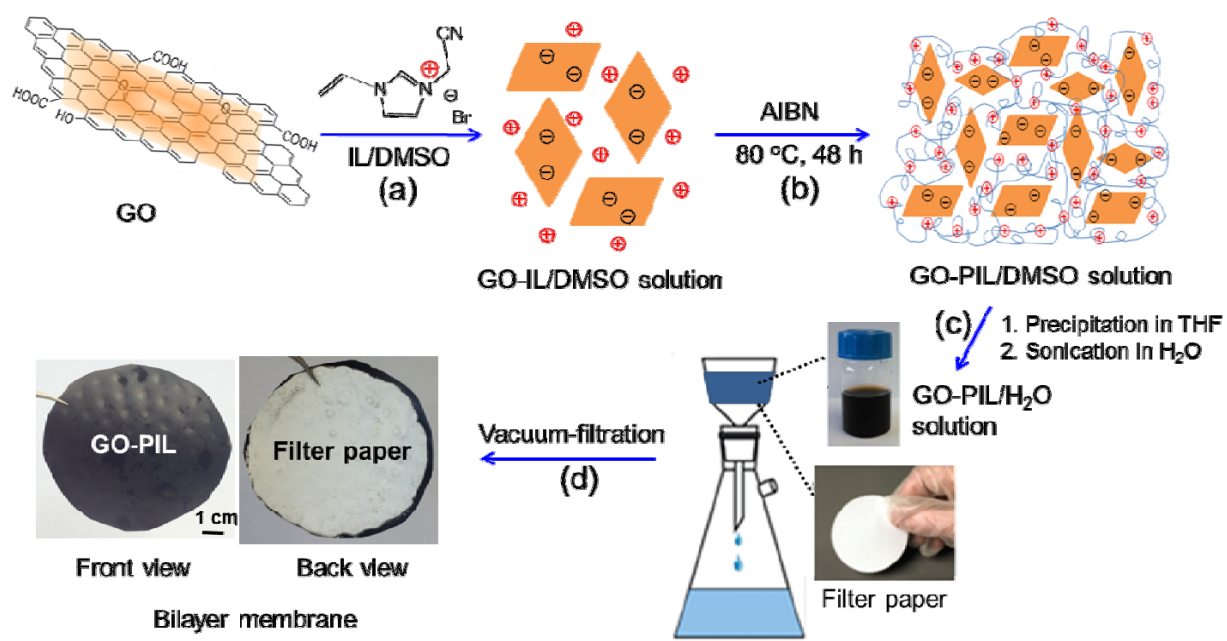

**Figure 1.** Schematic illustration of the preparation of a GO-PIL/filter paper bilayer membrane. (a) The dispersion of GO and IL in DMSO, (b) *in situ* polymerization of IL grafted on GO, (c) the collection of the GO-PIL product and preparation of GO-PIL/H$_2$O solution, (d) vacuum filtration to fabricate the GO-PIL/filter paper membrane.

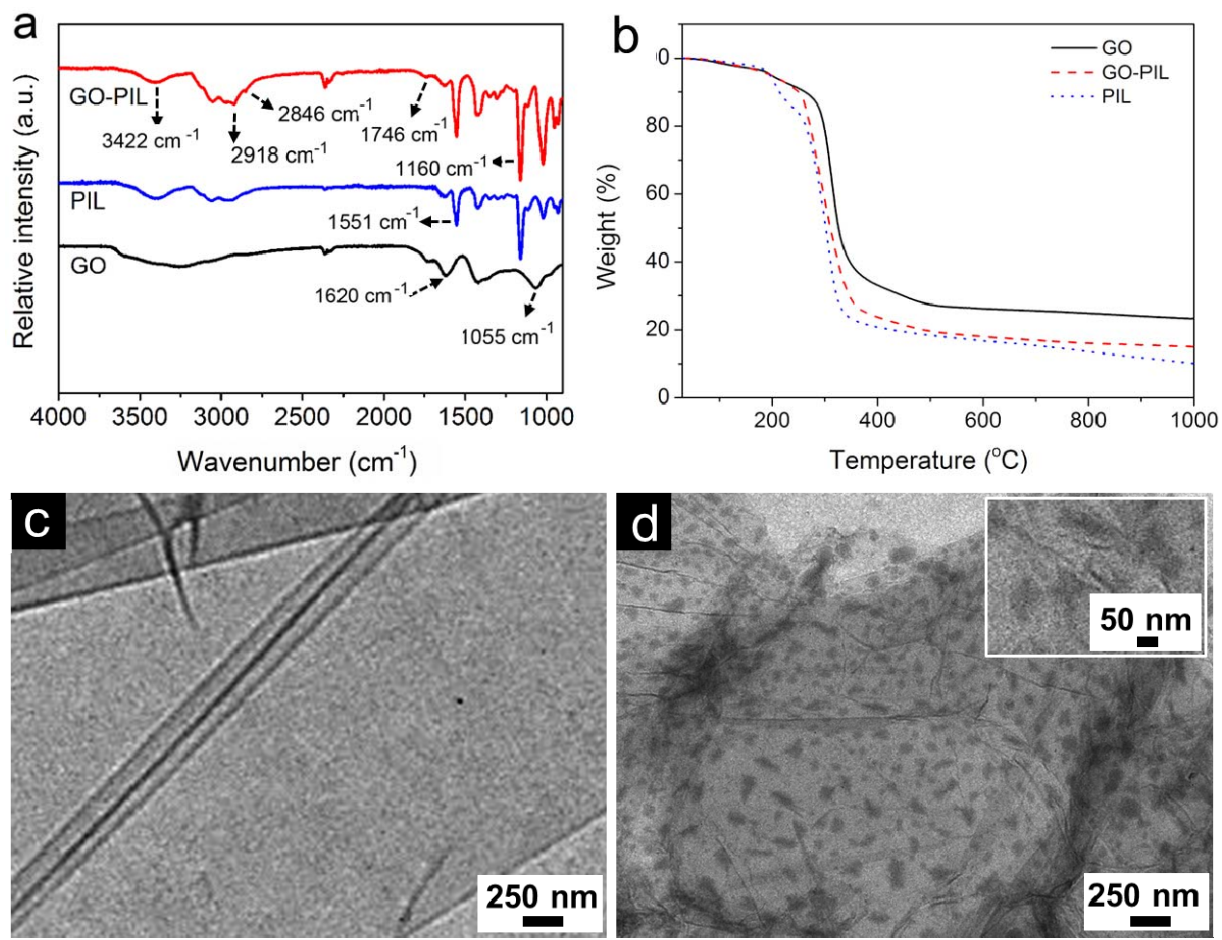

**Figure 2.** a) FT-IR spectra of GO, PIL and GO-PIL; b) TGA curves of GO, PIL and GO-PIL; c) TEM image of GO; d) TEM image of GO-PIL hybrid nanosheets in low and high magnifications (inset in d).

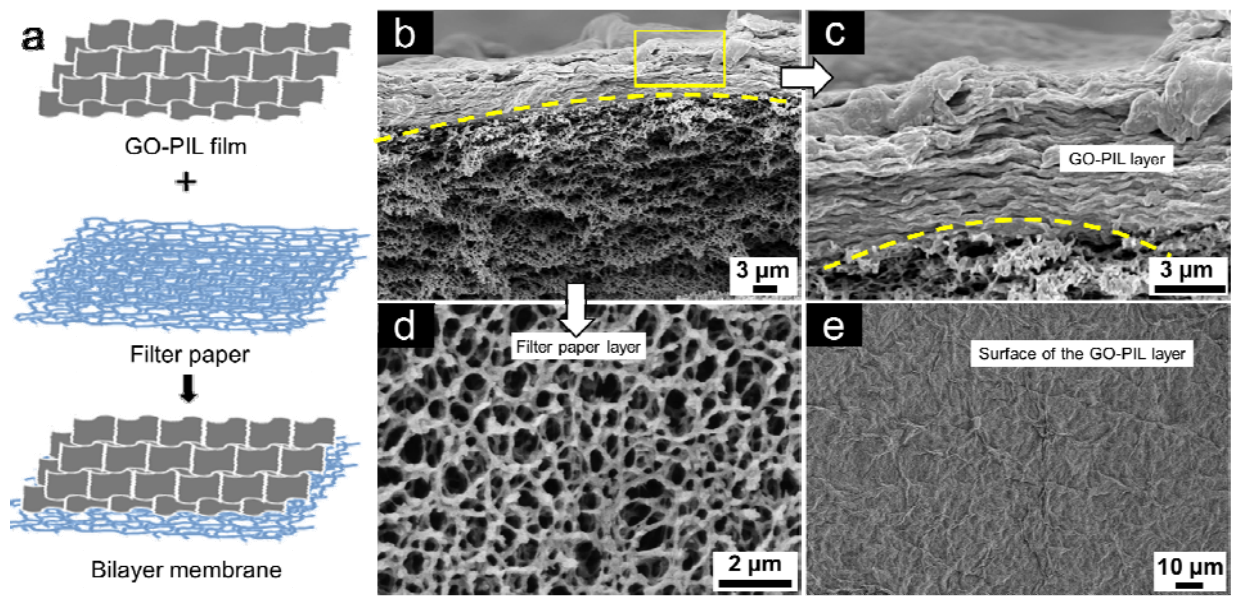

**Figure 3.** a) Schematic illustration of the bilayer structure of the GO-PIL@filter paper membrane. SEM images of: b) the cross-section of the GO-PIL/filter paper bilayer membrane; c) an enlarged view of the cross-section of the bilayer membrane, the GO-PIL film here is ca. 7 μm; d) the porous structure of the cellulose ester-based membrane filter paper; e) the top surface of the GO-PIL film.

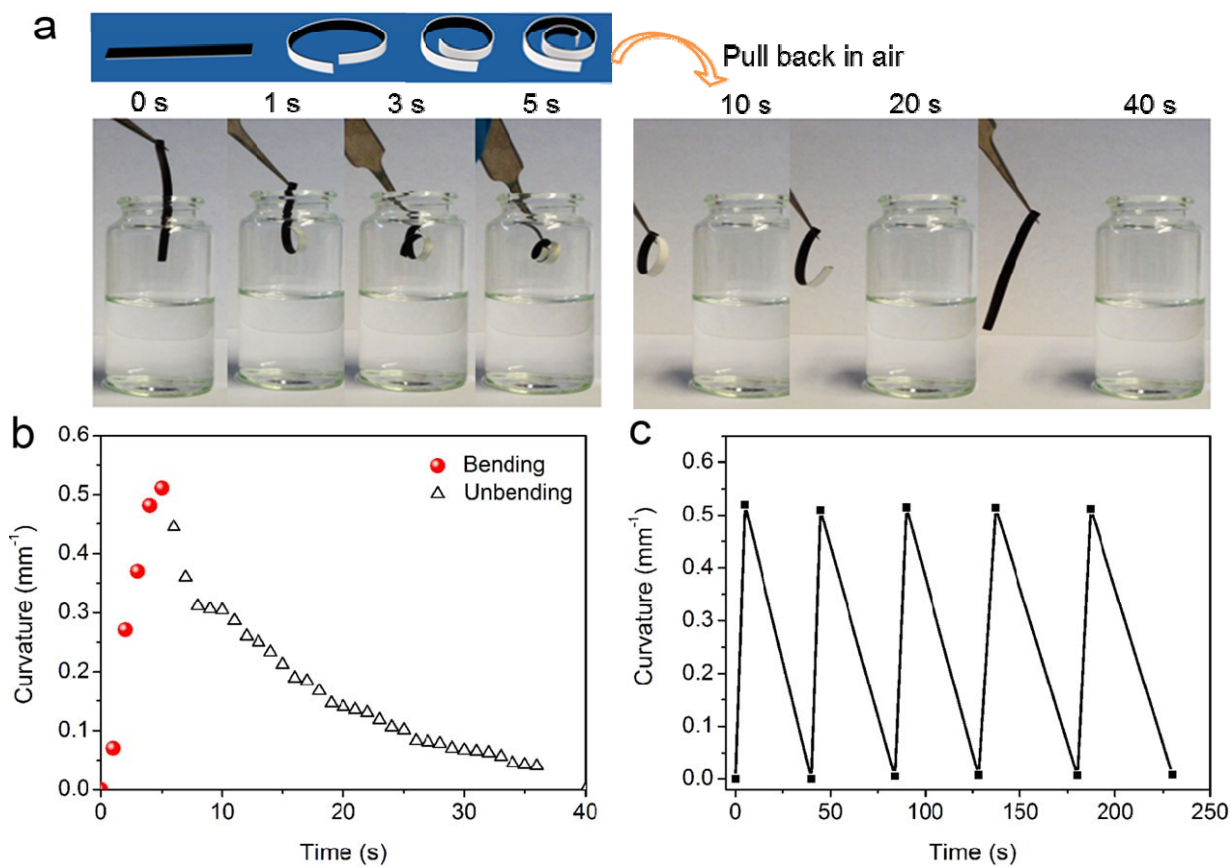

**Figure 4.** a) Adaptive movement of a GO-PIL/filter paper bilayer membrane (20 mm ×1 mm × 107 μm) placed in acetone vapor (20 °C) and then back in air; b) Plot of curvature *vs.* time for the membrane actuator in acetone vapor and then back in air; c) The reversible bending/unbending deformation of a GO-PIL/filter paper bilayer membrane actuator driven by periodic contact with acetone vapor (20 °C).

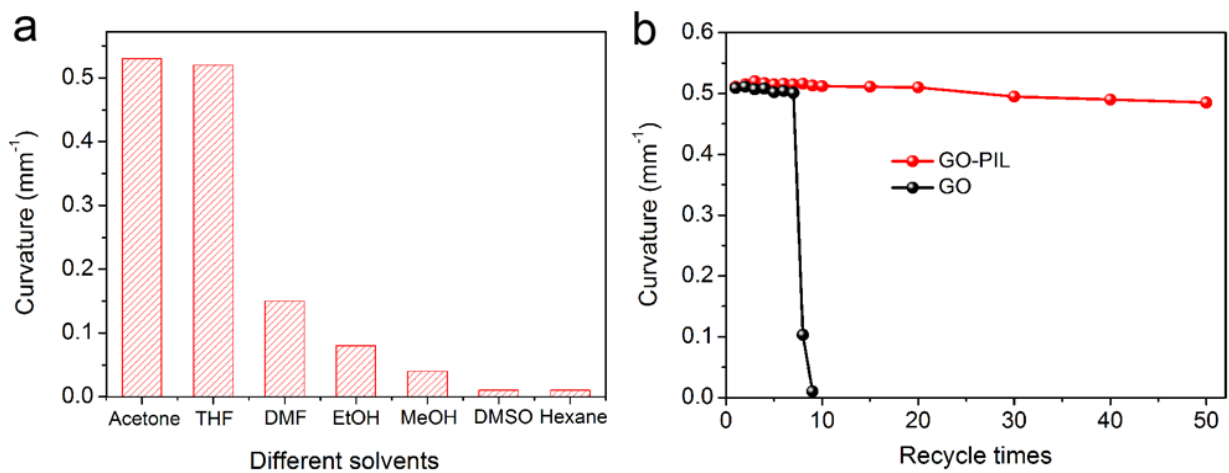

**Figure 5.** a) Curvatures of a GO-PIL/filter paper bilayer membrane in different solvent vapors (20 °C); b) Dependence of the curvature of the GO/filter paper and the GO-PIL/filter paper bilayer membrane on cycle times of the bending-unbending motion.

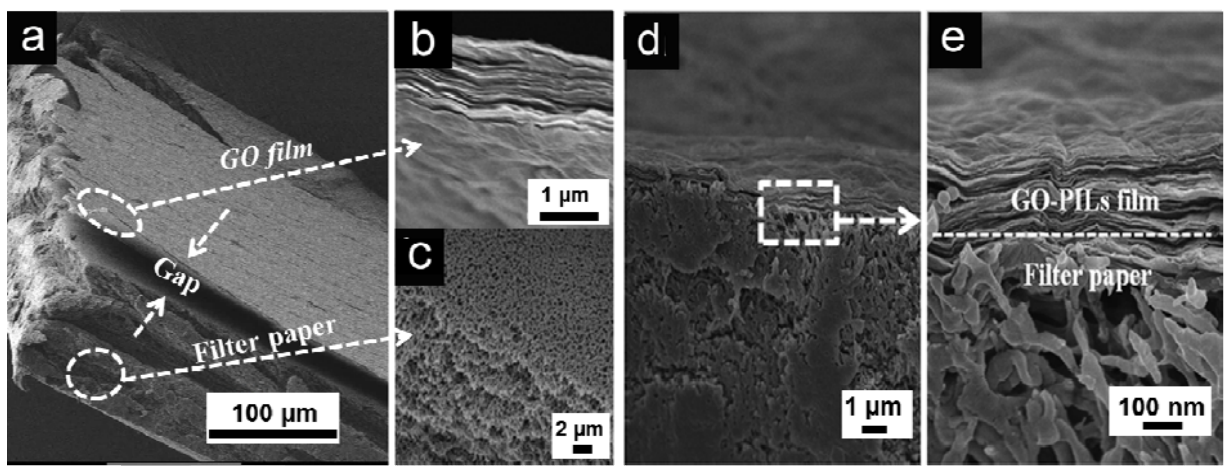

**Figure 6.** SEM images of the cross-section of the membranes after 8 cycles of bending-unbending process. a) a GO/filter paper membrane (the thickness ratio of GO to filter paper was 1:100), b) and c) a high magnification view of the GO film and the filter paper in a); d) the GO-PIL/filter paper membrane with 25.9 wt.% of PIL in the GO-PIL film (the thickness ratio of GO-PIL to filter paper was 1.5:100), e) a high magnification view of the GO-PIL film and the filter paper in d).

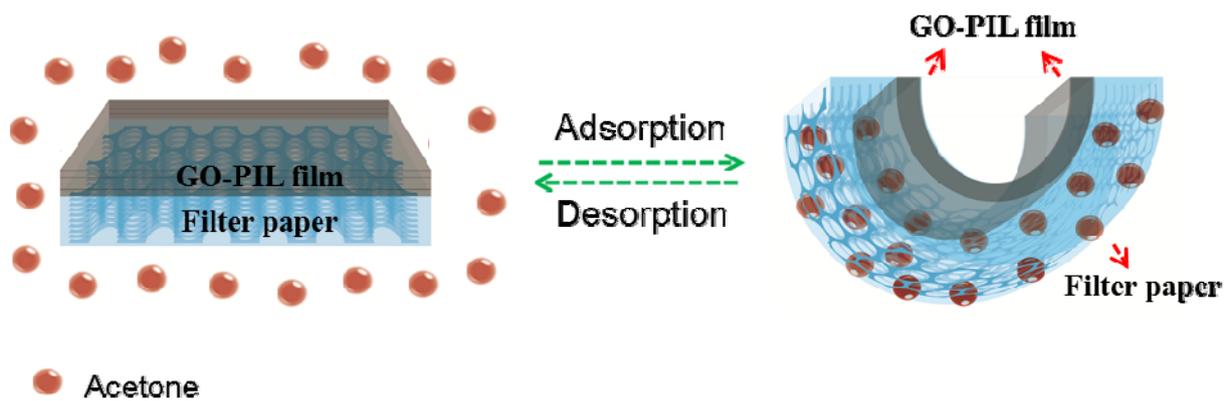

**Figure 7.** Schematic illustration of the bending-unbending actuation of a GO-PIL/filter paper bilayer membrane toward adsorption and desorption of acetone molecules.

**The Table of Contents**

**A porous bilayer membrane actuator** is fabricated from a porous filter paper deposited with a thin film of poly(ionic liquid)-modified graphene oxide nanosheets (GO-PIL). Due to the asymmetric swelling of the top GO-PIL film and the bottom porous filter paper towards organic vapor, the bilayer membrane shows rapid, reversible bending with top GO-PIL dense film inwards.

**Graphene oxide, poly(ionic liquid), filter paper, bilayer actuator**

H. Song, H. Lin, M. Antonietti, J. Yuan*

**From Filter Paper to Functional Actuator by Poly(ionic liquid)-Modified Graphene Oxide**

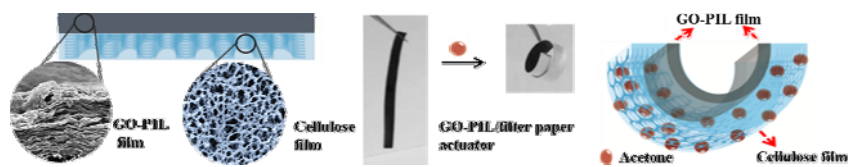

Supporting Information

**From Filter Paper to Functional Actuator by Poly(ionic liquid)-Modified Graphene Oxide**

*Haojie Song, Huijuan Lin, Markus Antonietti, Jiayin Yuan\**

**Descriptions of supporting movies**

Movie S1: The actuation behavior of the GO-PIL/filter paper membrane in acetone vapor.

Movie S2: The actuation behavior of the filter paper membrane in acetone vapor.